# Independent Tuning of Surface Acoustic-Waves and Spin-Waves via Buffer-Layer Engineering in $Co_2FeGe$ Heusler Thin Films

A. V. Achuthan[a*], A. Vovk[b], S. Bunyaev[b], B. Postolnyi[b], P. Štrichovanec[c], P. A. Algarabel[c], K. Załęski[d], J. P. Araujo[b], G. N. Kakazei[b], A. Trzaskowska[a]

[a] *Institute of Spintronics and Quantum Information, Faculty of Physics and Astronomy, Adam Mickiewicz University, 61-614 Poznan, Poland*
[b] *Instituto de Física dos Materiais Avançados, Nanotecnologia e Fotónica, Universidade do Porto, Rua do Campo Alegre 687, 4169-007 Porto, Portugal*
[c] *Instituto de Nanociencia y Materiales de Aragón, Universidad de Zaragoza-CSIC, Campus Río Ebro, 50018 Zaragoza, Spain*
[d] *NanoBiomedical Centre, Adam Mickiewicz University, 61-614 Poznan, Poland*
** Corresponding author: amrvac@amu.edu.pl*

Abstract

Understanding and controlling acoustic and spin-wave excitations in magnetic thin films is critical for the development of magnonic and spin-acoustic devices. We report on the use of Cr and W buffer layers to independently modify the acoustic and magnetic excitations in $Co_2FeGe$ full-Heusler thin films grown on MgO(001). Using Brillouin light scattering (BLS) spectroscopy and ferromagnetic resonance (FMR), we probed Rayleigh and Sezawa surface acoustic waves (SAWs) alongside Damon-Eshbach and perpendicular standing spin-wave (PSSW) modes. Our results show that acoustic dispersion depends strongly on the buffer material; W-buffered films exhibit a pronounced 16% reduction in Rayleigh SAW frequency compared to buffer-free films, primarily due to mass loading and acoustic impedance shifts. While the buffer layers significantly shift acoustic frequencies, they simultaneously modify the dynamic magnetic response (increasing spin-wave group velocity by ~34%) through different physical mechanisms. Finite-element simulations show excellent agreement with the experimental acoustic data. These findings demonstrate that buffer-layer engineering is an effective strategy for the independent tailoring of elastic and magnetic excitations, providing a versatile platform for hybrid spin-acoustic technologies.

## 1. INTRODUCTION

Heusler alloys, first reported in the early twentieth century [1], constitute a versatile class of intermetallic compounds whose electronic and magnetic properties can be tuned over a broad range. Owing to their ordered crystal structures and compositional flexibility, these materials can exhibit metallic, semiconducting, half-metallic, or superconducting behavior [2, 3, 4]. They are commonly grouped into full-Heusler ($X_2YZ$), half-Heusler (XYZ), and related structural variants. Notably, many Heusler alloys exhibit strong magnetism despite being composed of non-ferromagnetic elements, highlighting the key role of atomic ordering and electronic structure in determining their functional response.

From a structural perspective, Heusler alloys are characterized by pronounced chemical ordering, typically realized in cubic crystal structures, which strongly governs their electronic, magnetic, and transport properties. This class of materials has expanded beyond conventional full- and half-Heusler systems to include inverse, binary, and quaternary compounds, significantly increasing its compositional and functional diversity [5]. Such structural versatility enables deliberate tuning of magnetic, electronic, and thermal characteristics within a unified material platform. Their compatibility with semiconductor lattices, along with low magnetic damping and an adjustable electronic structure, further makes Heusler alloys highly attractive for thin-film heterostructures and nanoscale spintronic, magnonic, and thermoelectric devices [6].

From a technological perspective, Heusler alloys are of particular interest because their electronic and magnetic properties [7, 8] can be tuned over a wide range through controlled variations in composition and atomic ordering. This tunability - reflected in key parameters such as spin polarization, Curie temperature, magnetic anisotropy, band gap, and carrier concentration - forms the basis for their use in spintronics [9], as well as thermoelectric [10], sensing [11, 12], refrigeration [13], and shape-memory applications [14]. In particular, cobalt-based full-Heusler thin films such as $Co_2FeAl$ and $Co_2FeGe$ are widely considered for spintronic devices due to their high Curie temperatures and high spin polarization [6]. Moreover, the coexistence of magnetic excitations and surface acoustic waves in thin-film heterostructures makes these systems relevant for studies of coupled elastic and spin-dynamic phenomena.

Previous studies on Heusler thin films have been primarily focused on optimizing their magnetic properties through parameters such as annealing temperature [15], chemical composition [16], and film thickness [17]. In parallel, buffer-layer engineering has been shown to improve structural quality [18] and to tune magneto-elastic behavior [19]. Buffer layers may also influence magnetic damping [17,20]. However, a systematic comparison of how different buffer layers influence both surface acoustic wave (SAW) propagation and spin-wave (SW) dispersion in $Co_2FeGe$/MgO heterostructures is still lacking. As a result, the elastic properties of Heusler thin films and their relation to acoustic excitation dynamics remain less explored than their magnetic properties.

Brillouin light scattering (BLS) is particularly suitable for this purpose because it allows non-contact detection of thermally excited surface acoustic waves in the GHz frequency range. When combined with finite-element simulations, it enables the identification of Rayleigh and Sezawa branches and the evaluation of how acoustic modes are localized near the surface. In a non-transparent multilayer, interpreting BLS spectra requires care because the acoustic wave can penetrate deeper into the structure than the optical probing depth. Thus, comparison between the experiment and the simulation should account for surface localization and the effective acoustic response of the near-surface multilayer.

Buffer-layer engineering offers a practical route for modifying the elastic and magnetic boundary conditions of $Co_2FeGe$/MgO films. A Cr buffer layer may affect epitaxial quality and interfacial strain, whereas a W buffer layer is expected to produce a stronger modification of surface-acoustic-wave dispersion because of its high mass density and comparatively low transverse acoustic velocity relative to MgO. Since MgO has a higher transverse acoustic velocity than $Co_2FeGe$ and the metallic buffer layers, the studied structures can be considered slow-on-fast layered systems [21], in which the introduction of a slower and denser interlayer is expected to decrease the SAW branches and promote guided Sezawa-type modes. Despite the relevance of these effects, a direct comparison of how Cr and W buffer layers modify surface-acoustic-wave dispersion in $Co_2FeGe$/MgO, accompanied by spin-wave characterization, has not yet been established.

Therefore, we treat the buffer layer primarily as a parameter controlling the effective acoustic response of the $Co_2FeGe$/MgO heterostructure. In parallel, the spin-wave spectra are analyzed to determine whether the same buffer layers also modify the effective magnetic response of the $Co_2FeGe$ film.

Here, we investigate $Co_2FeGe$ Heusler thin films grown on MgO(001) substrates with three buffer-layer configurations: no buffer layer, a Cr buffer layer, and a W buffer layer.

Brillouin light scattering is used to measure Rayleigh and Sezawa surface acoustic-wave modes, as well as Damon–Eshbach (DE) and perpendicular standing spin-wave (PSSW) modes. Finite-element simulations are employed to interpret the acoustic dispersion, while ferromagnetic resonance provides complementary information on the effective magnetic parameters. The main aim of this work is to determine how the buffer-layer material tunes the SAW dispersion in $Co_2FeGe$/MgO and how this correlates with shifts in the SW resonance frequencies and mode structure.

## 2. MATERIALS AND METHODS

### 2.1. Sample preparation

Heusler alloy $Co_2FeGe$ films of 50 nm thickness were grown on 10×10 mm$^2$ MgO (001) substrates by DC magnetron co-sputtering technique using Orion-5 deposition system (by AJA International Inc.). The system was equipped with a cluster assembly of 5 confocal magnetron sources. Deposition was performed from two independent direct current (DC) 2" magnetrons loaded with high purity (better than 99.995 at. %) $Co_2Fe$ alloy and Ge targets. Base pressure was below $2\times10^{-7}$ Torr, and the depositions were carried out at 3 mTorr of Ar. The deposition rates were estimated from reference films' thickness measurements. The values of 0.05 nm/sec for $Co_2Fe$ and 0.032 nm/sec for Ge were used to ensure stoichiometric composition of the resulting alloy.

For seed (buffer) layers, high-purity (purity exceeding 99.995 %) Cr and W targets were used. They were mounted on separate DC magnetron sources of the cluster assembly. The buffer layers were approximately 15 nm thick. The deposition rates were set at 0.043 nm/sec and 0.05 nm/sec for Cr and W, respectively. During film preparation, the deposition rates for all magnetrons were kept constant via computer control of the current, allowing reproduction of the technological conditions with high accuracy. The substrates were mounted on a rotating sample holder (30 RPM) that was positioned 120 mm from the deposition sources. All films were deposited at room temperature (RT, 293 K). The schematic diagram of the studied $Co_2FeGe$ with different buffer layers is shown in Fig. 1.

The composition of the films was determined by Energy Dispersive X-Ray analysis (FEI Quanta 400FEG Field Emission SEM/EDAX-PEGASUS X4M) as $Co_{48}Fe_{24}Ge_{28}$ (±2 at. %) that agrees well with the nominal stoichiometric composition.

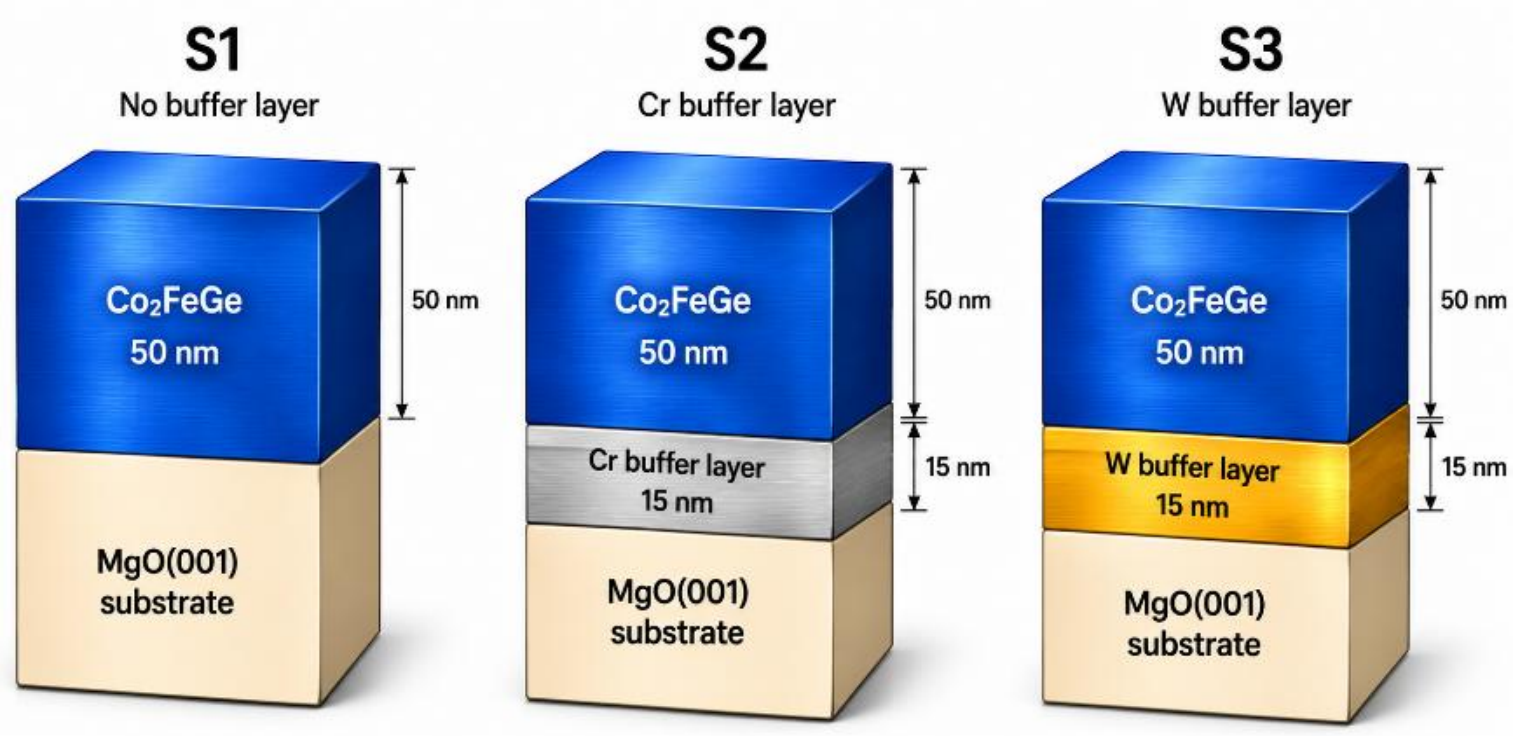

*Fig. 1: Schematic illustration of the investigated $Co_2FeGe$/MgO heterostructures with different buffer-layer configurations: (S1) buffer-free $Co_2FeGe$, (S2) $Co_2FeGe$ /Cr, and (S3) $Co_2FeGe$ /W. All films have a $Co_2FeGe$ thickness of approximately 50nm and buffer layers of 15nm where applicable.*

## 2.2. Experimental Methods

### 2.2.1. Structural characterization

Grazing-incidence X-ray diffraction (GIXRD) in parallel-beam, out-of-plane geometry and X-ray reflectivity (XRR) were performed using a Rigaku SmartLab diffractometer equipped with a primary Ge monochromator (Cu-Kα radiation). The GIXRD configuration restricts the beam's penetration depth, enhancing sensitivity to the $Co_2FeGe$ films and buffer layers and minimizing substrate contribution [22]. XRR profiles were fitted using LEPTOS software v2.02 to evaluate thicknesses of the film and the native oxide formed on the surface, as well as interfacial and surface roughness.

### 2.2.2. Magnetostatic Measurements

Magnetostatic measurements were performed using a Quantum Design MPMS-XL SQUID magnetometer. The samples were measured at RT in a field range of ±10 kOe. Because the investigated samples consist of $Co_2FeGe$ films deposited on different buffer layers, the measured SQUID signal includes both the ferromagnetic contribution from the $Co_2FeGe$ layer and background contributions from the substrate and non-magnetic layers in the stack. To isolate the magnetic response of $Co_2FeGe$, a linear background was fitted to the high-field saturated region and subtracted from the measured magnetic moment. The saturation magnetization ($M_S$) was subsequently determined by normalizing the corrected saturation moment to the volume of the $Co_2FeGe$ layer [24,24].

### 2.2.3 Ferromagnetic Resonance

Broadband ferromagnetic resonance (FMR) measurements were conducted at RT using a custom-built setup that included a gold coplanar waveguide (CPW) connected to an Anritsu 37247D vector network analyzer (VNA). A DC magnetic field was applied in the film plane perpendicularly to the direction of the magnetic component of the microwave field. The transmission parameter $S_{21}$ was measured as a function of the external magnetic field in a frequency range up to 20 GHz. Detailed descriptions of the FMR experimental and fitting procedures were reported in Refs. [25, 26].

The resonance condition was analyzed using the Kittel relation for thin in-plane magnetized ferromagnetic films [27]:

$$f = \left(\frac{\gamma}{2\pi}\right)\sqrt{(H_{ext} + H_{ani})(H_{ext} + H_{ani} + 4\pi M_{eff})} \quad (1)$$

where $H_{ani}$ is the anisotropic field, $\gamma$ is the gyromagnetic ratio, and $M_{eff}$ is the effective magnetization. $M_{eff}$ is related to the saturation magnetization $M_S$ and perpendicular anisotropy field $H^{\perp}_{ani}$ through:

$$4\pi M_{eff} = 4\pi M_s - H^{\perp}_{ani} \quad (2)$$

Analysis of the resonance spectra enabled the determination of the effective magnetization and magnetic anisotropy of the $Co_2FeGe$ thin films.

To account for PSSW modes, the exchange contribution was included by introducing a quantized wavevector $k_p = p\pi/h$, where $p$ is the mode number and $h$ is the film thickness. In this case, the resonance frequency is described by the extended Kittel relation:

$$f_{PSSW} = \left(\frac{\gamma}{2\pi}\right)\sqrt{\left(H_{ext} + H_{ani} + \left(\frac{2A_{ex}}{M_{eff}}\right)\left(\frac{p\pi}{h}\right)^2\right)\left(H_{ext} + H_{ani} + 4\pi M_{eff} + \left(\frac{2A_{ex}}{M_{eff}}\right)\left(\frac{p\pi}{h}\right)^2\right)} \quad (3)$$

where $A_{ex}$ is the exchange stiffness constant and the term $(2A_{ex}/M_{eff})(p\pi/h)^2$ represents the exchange field contribution associated with the quantized spin-wave modes, and $\gamma/2\pi$ is the gyromagnetic ratio, which describes the precessional response of the magnetization to an applied magnetic field. This expression was used to fit the observed PSSW resonances and extract the exchange stiffness and mode indices. The gyromagnetic ratio $\gamma$ is given by:

$$\gamma = \frac{g\mu_B}{\hbar} \quad (4)$$

where g is the Landé g-factor, $\mu_B$ is the Bohr magneton, and $\hbar$ is the reduced Planck constant.

### 2.2.4 Brillouin Light Scattering

Brillouin light scattering spectroscopy was used to probe thermally excited phonon and magnon modes in the studied $Co_2FeGe$ thin films. This technique enables the detection of both surface acoustic waves and spin-wave modes without external excitation, providing direct access to their frequencies and dispersion characteristics. In the phonon spectra, Rayleigh and Sezawa surface acoustic modes can be resolved, whereas the magnon spectra allow the identification of Damon-Eshbach and perpendicular standing spin-wave modes.

The BLS spectra were acquired at room temperature using a six-pass tandem Fabry-Perot interferometer from JRS Scientific Instruments. A continuous-wave laser with a wavelength of $\lambda_0 = 532$ nm was used as the light source. More information can be found in Ref. [28]. The incident laser power was limited to 100 mW to minimize laser-induced heating while maintaining a sufficient signal-to-noise ratio. For phonon measurements, incident and scattered light were *p*-polarized in the sagittal scattering plane [29]. For spin-wave measurements, a cross-polarized *p–s* geometry was used [30].

The measurements were performed in the backscattering geometry, in which the transferred in-plane wavevector $q$ is given by the momentum conservation condition:

$$q = \frac{(4\pi sin\theta)}{\lambda_0} \quad (5)$$

where $\theta$ is the angle between the incident light and the sample surface normal, and $\lambda_0$ is the wavelength of the incident laser light. The acoustic phase velocity was calculated from the measured Brillouin frequency shift as:

$$v_{ph} = \frac{(\Delta f \cdot \lambda_0)}{2sin\theta} \quad (6)$$

By varying the angle of incidence, the transferred wavevector was tuned from 6 $\mu m^{-1}$ to 24 $\mu m^{-1}$, enabling the determination of the phonon dispersion relations.

For the spin-wave measurements, an external magnetic field was applied in the plane of the sample to define the equilibrium magnetization direction and control the magnon spectrum. The measurements were carried out in the Damon-Eshbach geometry, where the spin-wave wavevector is perpendicular to the applied magnetic field. This configuration allows the detection of surface-localized Damon-Eshbach modes as well as perpendicular standing spin-wave modes. The magnetic origin of the observed peaks was verified through field-dependent BLS measurements performed between 500 and 3500 Oe.

Each spectrum was accumulated for 30 min and fitted using Lorentzian line shapes to determine the peak positions. The uncertainties in the extracted frequencies and transferred wavevectors were approximately 0.07 GHz and 0.001 $\mu m^{-1}$, respectively.

### 2.3. Finite-Element Simulations

Numerical simulations of acoustic phonon propagation were performed using the finite-element method (FEM) implemented in COMSOL Multiphysics [31, 32]. The geometric model explicitly included the MgO substrate, the buffer layer, and the $Co_2FeGe$ thin film, each treated as a separate layer with their respective material parameters. The material parameters used in the simulations, including density and elastic constants, are summarized in Table 1.

Table 1: Elastic constants ($c_{11}$, $c_{12}$, $c_{44}$) and density ($\rho$) of the substrate MgO, buffer materials Cr and W and the studied samples.

| Layer | $c_{11}$ (GPa) | $c_{12}$ (GPa) | $c_{44}$ (GPa) | $\rho$ (kg/m$^3$) |
|---|---|---|---|---|
| MgO [33] | 295.9 | 95.4 | 153.9 | 3585 |
| Cr [34] | 339.8 | 57.8 | 100.8 | 7200 |
| W [35] | 523.5 | 204.9 | 159 | 19300 |
| $Co_2FeGe$ [36] | 229.2 | 160.3 | 123.4 | 7749 |

Since an experimental density value for the present $Co_2FeGe$ thin films was not available, the $Co_2FeGe$ density was estimated using a composition-weighted rule-of-mixtures approach:

$$\rho(Co_2FeGe) = 0.50\,\rho(Co) + 0.25\,\rho(Fe) + 0.25\,\rho(Ge) \quad (7)$$

The MgO substrate thickness in the simulation domain was set to 20 µm. Bloch-Floquet periodic boundary conditions were applied along the in-plane directions. The lattice displacement field u(r, t) associated with the acoustic vibration was expressed as a plane-wave solution:

$$u(r,t) = u_0 exp[i(q \cdot r - \varpi t)] \quad (8)$$

The phonon dispersion relations were obtained by solving the elastic wave equation in the frequency domain for different in-plane wavevectors. A non-uniform mesh was employed, with a finer mesh applied near the film surface. Mesh-size and substrate-thickness convergence tests were performed to ensure that the calculated SAW frequencies changed by less than 2% after further mesh refinement. Detailed descriptions of the simulations can be found in [28,37].

3. RESULTS AND DISCUSSION

*3.1 X-ray analysis (GIXRD and XRR)*

Only (022) and (004) $Co_2FeGe$ reflections were observed for all films, indicating polycrystalline growth. No superlattice reflections characteristic of B2 and/or $L2_1$ phases were detected, pointing to the formation of an atomically disordered A2 structure of $Co_2FeGe$. The low adatom mobility at room temperature and relatively low deposition rates both lead to the suppression of epitaxial growth and to the formation of an A2 disordered atomic structure. Out-of-plane lattice parameter (a) determined from the (004) peak is a = 5.712 Å (S1), 5.690 Å (S2), and 5.714 Å (S3). These values are slightly below bulk $Co_2FeGe$ (5.738 Å) but consistent with those previously reported for films grown under similar conditions [25]. The Cr seed-layer structure could not be resolved because Cr (011) and (002) reflections overlap $Co_2FeGe$ (022) and (004), respectively. In contrast, the W buffer shows a mixture of α-W (stable A2) and β-W (metastable A15) phases. This is a typical result for W films sputtered at room temperature [38]. For S1 and S3, fits of XRR profiles return a $Co_2FeGe$ thickness of ≈47 nm with ≈1.5 nm native surface oxide. The combined thickness corresponds well to the targeted value of 50 nm. The W buffer in S3 was estimated to be 15.8 nm thick. Surface roughness is ≈1.5 nm (S1) and ≈1.3 nm (S3); the W buffer roughness is <0.3 nm. No satisfactory fit could be obtained for sample S2. This is most likely due to the limitations of XRR, which is best suited for continuous films with a laterally homogeneous thickness over the probed area [39]. A possible explanation is that, under the given deposition conditions, Cr exhibits Volmer–Weber (island-type) growth. Consequently, the $Co_2FeGe$ film grows on a corrugated surface formed by Cr islands, resulting in increased surface roughness. Another explanation is the diffusion of Cr into the $Co_2FeGe$ layer, leading to the formation of a broadened interfacial region. Such a mechanism was reported in [40] for Co-based Heusler alloy films annealed at high temperatures. However, this explanation can be ruled out in the present case because no post-deposition heat treatment was performed.

*3.2 Magnetostatic properties (SQUID Magnetometry)*

Hysteresis loops for samples S1, S2, and S3 are shown in Figure 2. All three samples exhibit ferromagnetic behavior with narrow hysteresis loops. The values of $M_S$ derived from these measurements are presented in Table 2. These are somewhat lower compared to 1200 emu/cm$^3$ found in bulk $Co_2FeGe$ alloy [41] and foils [42]. However, they are within the range reported earlier for films deposited in similar conditions [25,26]. The reduction in $M_S$ can be due to a slight Ge enrichment in the films and atomic disorder. Among the samples investigated, S3 exhibits the highest saturation magnetization, followed by S1, whereas S2 shows the lowest value. The values of coercive fields ($H_c$) were determined as 11 Oe (S1), 32 Oe (S2), and 12 Oe (S3). The reduced $M_S$ and increased $H_c$ for sample S2 might be due to the rough microstructure of the film with a higher density of pinning centers. A similar increase in $H_c$ was reported earlier for polycrystalline films of $Co_2FeGe$ with enhanced surface roughness [25].

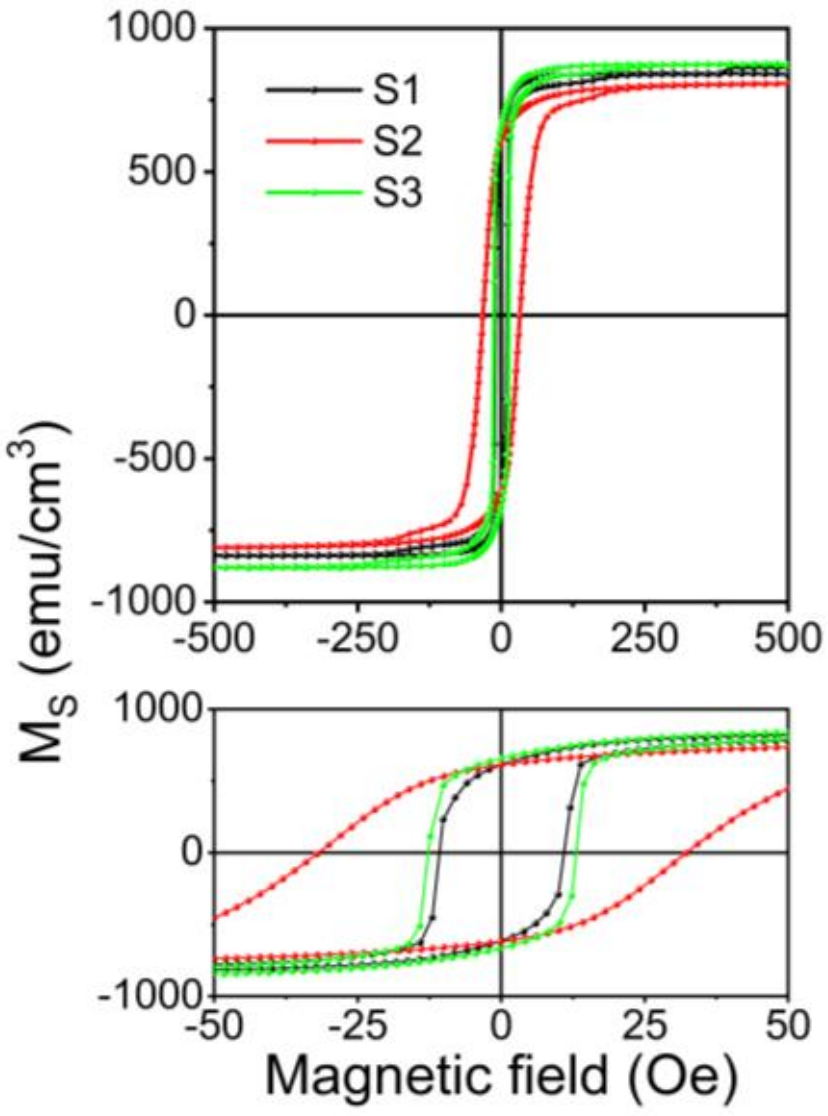

*Fig. 2: Magnetic hysteresis loops for samples: S1(black), S2 (red) and S3 (green) The saturation magnetization values determined from SQUID measurements are presented in Table 2.*

*3.3 Magnetodynamic Properties (VNA-FMR and Field-Dependent BLS)*

The results of VNA-FMR measurements, presented as frequency vs. magnetic field (H) maps, are shown in Fig. 3. For samples S1 and S3, two resonance lines were observed. They were identified as uniform FMR precession and the first PSSW modes. Solid lines drawn over experimental results represent fit of the data for $f_{FMR}$ and $f_{PSSW}$ using Eqs. (1) and (3), respectively. Estimated values of $M_{eff}$, gyromagnetic ratio ($\gamma/2\pi$), and exchange stiffness ($A_{ex}$) are presented in Table 2. The in-plane anisotropy $H_{ani}$ extracted using Eq. (1) was found to be vanishing (< 15 Oe for all cases). This is a typical value for fine polycrystalline films due to the randomization of the magnetocrystalline anisotropy axis within the sample volume. It provides additional indirect evidence for the assumption that epitaxy was suppressed and polycrystalline films were grown. The estimated value for gyromagnetic ratio $\gamma/2\pi$ is ≈2.86

MHz/Oe. It corresponds to the g-factor of ≈2.04, which indicates weak spin-orbit interactions. This result correlates well with the earlier reports on $Co_2FeGe$ films of different compositions [43]. Consistent with the structural investigations, sample S2 exhibited a broader FMR linewidth compared to the other samples, which is likely associated with extrinsic relaxation mechanisms such as inhomogeneous broadening induced by the rougher $Co_2FeGe$/Cr interface. Also, the intensity of the 1st PSSW for sample S2 was low and beyond the sensitivity of our experimental setup. This prevented an estimation of $A_{ex}$ for sample S2 from VNA-FMR measurements.

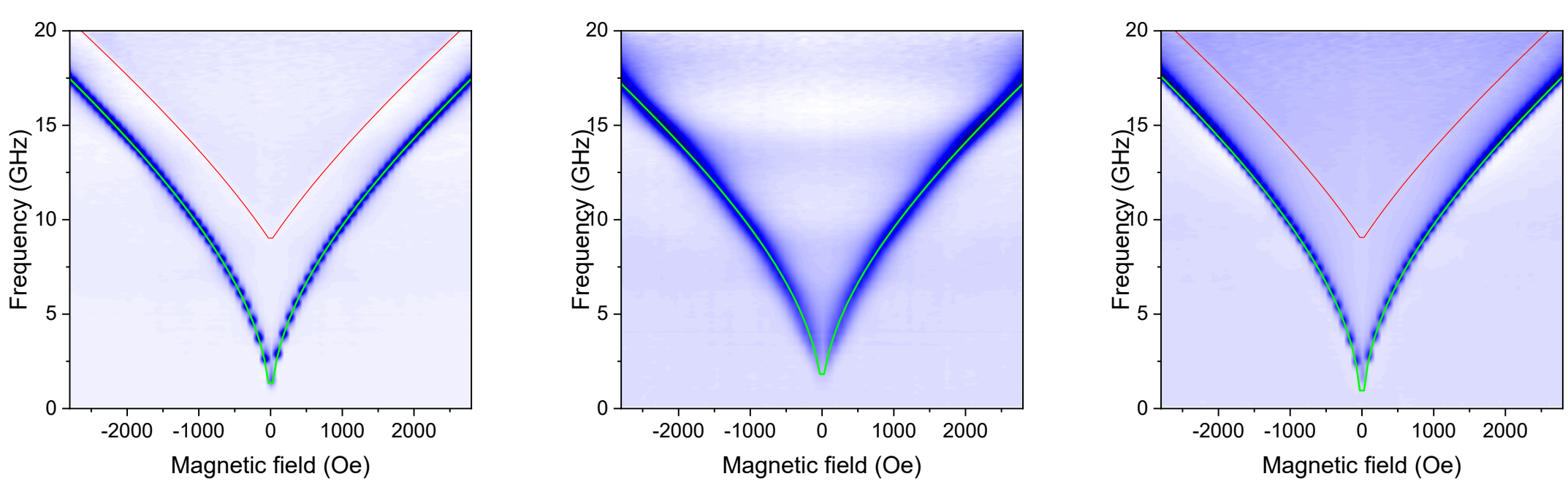

*Fig. 3: VNA-FMR frequency vs. field maps for samples S1 (a), S2 (b), and S3 (c). The solid lines show the results of fittings for the uniform FMR mode (green) and the 1st PSSW (red). The derived values of ($M_{eff}$), gyromagnetic ratio ($\gamma/2\pi$), and exchange stiffness ($A_{ex}$) are presented in Table 2.*

The magnon excitations of all investigated samples were measured by BLS spectroscopy (see Fig. 4). In the magnon spectra, the Damon-Eshbach (D-E) mode and perpendicular standing spin-wave (PSSW) modes are observed for all samples studied. In contrast to the phonon peaks, the Stokes and anti-Stokes intensities of the spin-wave modes are asymmetric, which is typical of magneto-optical light scattering from magnetic thin films [44,45]. The magnetic origin of these modes is confirmed by field-dependent measurements, in which the corresponding peaks shift toward higher frequencies as the external magnetic field increases. By contrast, the phonon peaks remain insensitive to the magnetic field, enabling a clear distinction between elastic and magnetic excitations [46].

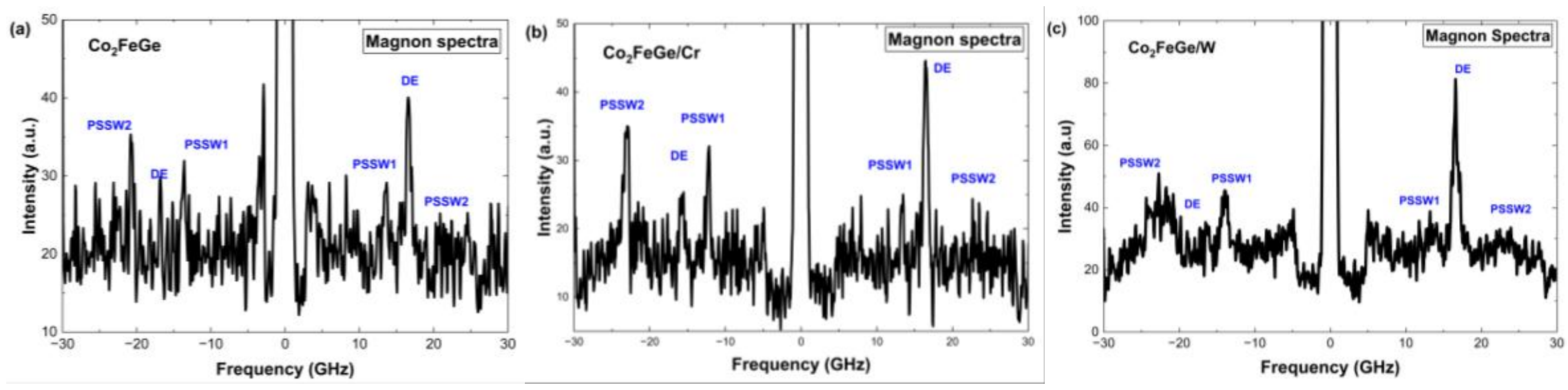

*Fig. 4: BLS spectra of magnon modes for the samples S1 (a), S2 (b) and S3 (c) at q = 13.1 $\mu m^{-1}$. The Damon–Eshbach (DE) and perpendicular standing spin-wave (PSSW) modes are observed.*

The evolution of the BLS spectra with applied magnetic field is shown in Fig. 5. The observed peaks shift toward higher frequencies with increasing in-plane magnetic field, confirming their magnetic origin. This behavior is absent for the acoustic modes, whose frequencies remain field independent. The lower-frequency magnetic branch is assigned to the Damon-Eshbach mode, whereas the higher-frequency branch is attributed to a perpendicular standing spin-wave mode.

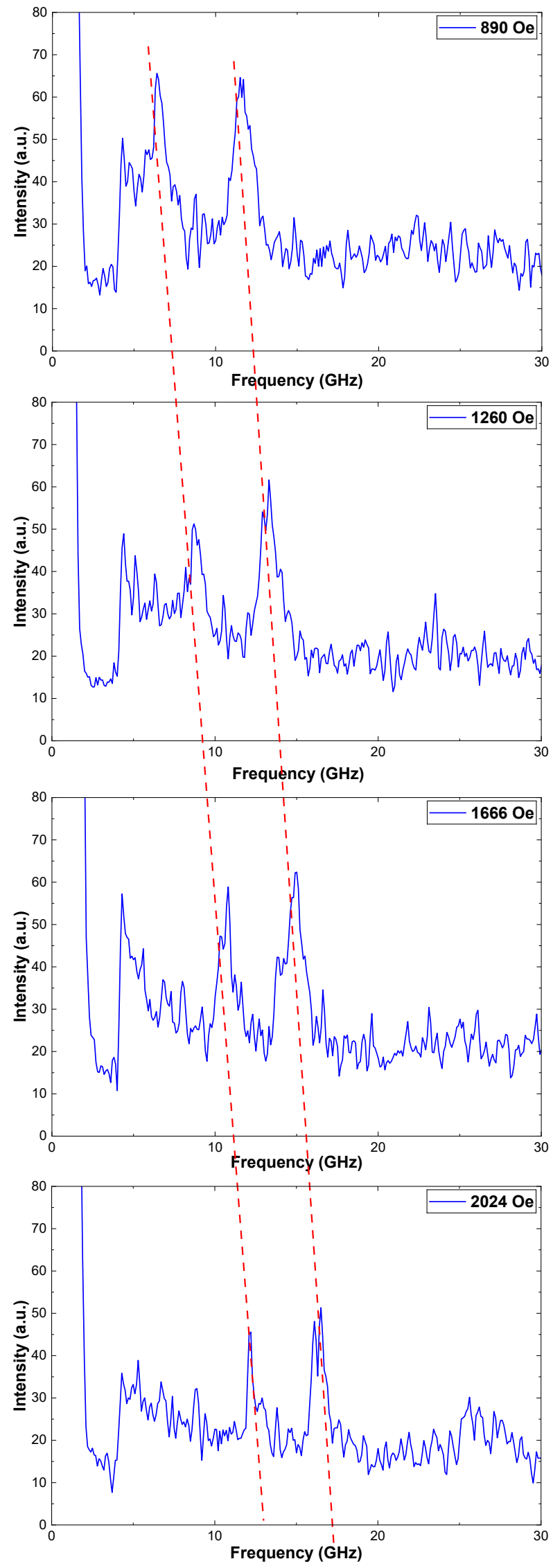

*Fig. 5: Magnetic-field dependence of the BLS magnon spectra for S1 measured at wavevector q≈0. The shift of the spectral peaks toward higher frequencies with increasing in-plane magnetic field confirms their magnetic origin.*

The field-dependent BLS measurements for all three samples are shown in Fig. 6. The BLS spectra were recorded by varying the in-plane magnetic field from 500 to 3600 Oe for *q≈0*. Two magnetic branches are observed: the lower-frequency branch is assigned to the Damon–Eshbach mode, whereas the higher-frequency branch is assigned to a PSSW mode. The experimental data were fitted simultaneously using the Kittel equation for the DE mode and the exchange-modified Kittel model for the PSSW mode, and values of $\gamma/2\pi$ and $M_{eff}$ are summarized in the Figures and presented in Table 2. It is worth noting an extremely good correlation between γ/2π and $M_{eff}$ determined by VNA-FMR and BLS measurements. Although the two techniques rely on different excitation and detection mechanisms and probe different spin-wave wavevector regimes, they yield nearly identical values of the gyromagnetic ratio and effective magnetization. This excellent agreement confirms the complementary nature of VNA-FMR and BLS and demonstrates the reliability of the determined magnetodynamic parameters.

The values of $A_{ex}$ allow evaluation of the exchange length $l_{ex}$. This parameter governs the characteristic length scale of magnetization reversal and is therefore important for both practical applications and micromagnetic simulations. Physically, $l_{ex}$ represents the characteristic distance over which a perturbation induced by the reversal of a single spin decays in a soft magnetic material. In [47] $l_{ex}$ was expressed as $l_{ex} = \sqrt{A_{ex}/2\pi M_S^2}$. Using the values listed in Table 2, $l_{ex}$ is estimated to be approximately 4.2 nm, 4.6 nm, and 4.7 nm for samples S1, S2, and S3, respectively. These values are typical of soft magnetic materials [47] and are comparable to those reported for polycrystalline $Co_2FeGe$ films [25].

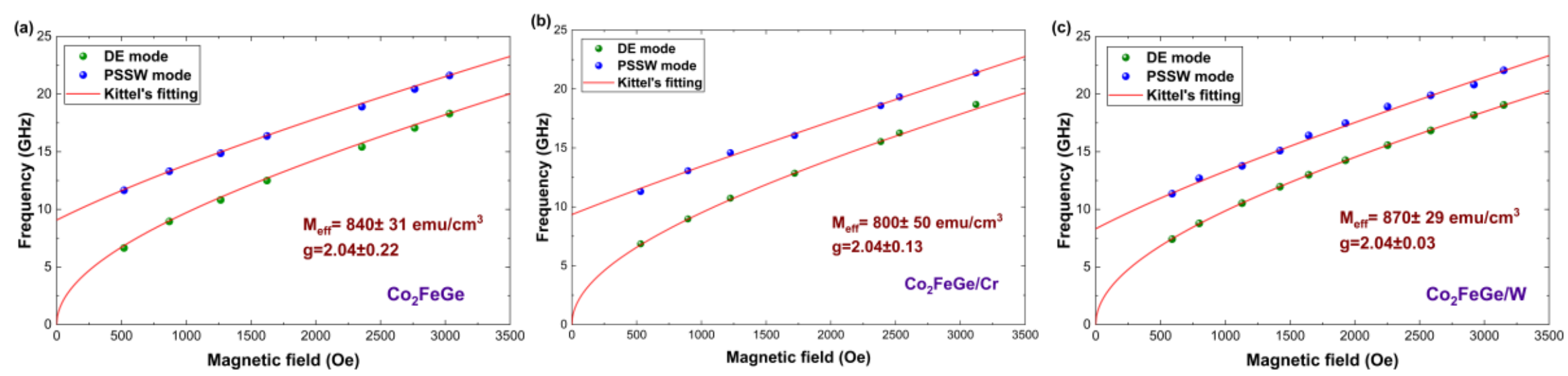

*Fig. 6: Field dependence of studied samples S1 (a), S2 (b), and S3 (c), respectively. The experimental data points correspond to the measured field-dependent resonance frequencies, while the solid lines show fits using the Kittel relation for the DE mode and the extended expression for the PSSW mode. The fitted values of Lande g-factor (g) and effective magnetization $M_{eff}$ are written inside the figures.*

Table 2. Comparison of magnetic parameters obtained from BLS, SQUID, and VNA-FMR measurements. Saturation magnetization ($M_S$) was determined from SQUID measurements. The effective magnetization ($M_{eff}$), gyromagnetic ratio ($\gamma/2\pi$), and exchange stiffness ($A_{ex}$), were extracted from spin-wave and ferromagnetic resonance analyses. The experimental error for $M_S$ was estimated from the uncertainty of the sample size determination. Error margins for BLS and VNA-FMR measurements were derived from a fitting procedure.

| Sample | BLS<br>$M_{eff}$ (emu/cm³) | BLS<br>$\gamma/2\pi$ (MHz/Oe) | BLS<br>$A_{ex}$ (erg/cm) | SQUID<br>$M_s$ (emu/cm³) |
|---|---|---|---|---|
| S1 | 840 ± 31 | 2.86 ± 0.31 | $9.56 \times 10^{-7}$ | 840± 20 |
| S2 | 800 ± 50 | 2.86 ± 0.18 | $8.49 \times 10^{-7}$ | 810± 20 |
| S3 | 870 ± 29 | 2.86 ± 0.04 | $8.17 \times 10^{-7}$ | 870± 20 |
| | | | | |
| | VNA-FMR<br>$M_{eff}$ (emu/cm³) | VNA-FMR<br>$\gamma/2\pi$ (MHz/Oe) | VNA-FMR<br>$A_{ex}$ (erg/cm) | |
| S1 | 833 ± 32 | 2.86 ± 0.04 | $9.50 \times 10^{-7}$ | |
| S2 | 809 ± 50 | 2.87 ± 0.01 | — | |
| S3 | 863± 59 | 2.85 ± 0.03 | $9.25 \times 10^{-7}$ | |

*3.4 Phonons investigations (BLS)*

The phonon BLS spectra reveal elastic excitations, enabling simultaneous investigation of the surface acoustic-wave dynamics (Fig. 7). In phonon spectra, the *Rayleigh* SAW (R-SAW) mode appears as the lower-frequency branch, whereas the *Sezawa* SAW (S-SAW) mode is observed at higher frequencies, which is characteristic of layered systems with acoustic-velocity contrast between the film and the substrate. Both modes exhibit systematic shifts depending on the buffer-layer material, indicating that the buffer layer modifies the effective elastic response of the bilayer system. Compared with the buffer-free sample, the Rayleigh mode shifts to lower frequencies for both buffered structures, with the largest reduction observed for the W-buffered (16%) film.

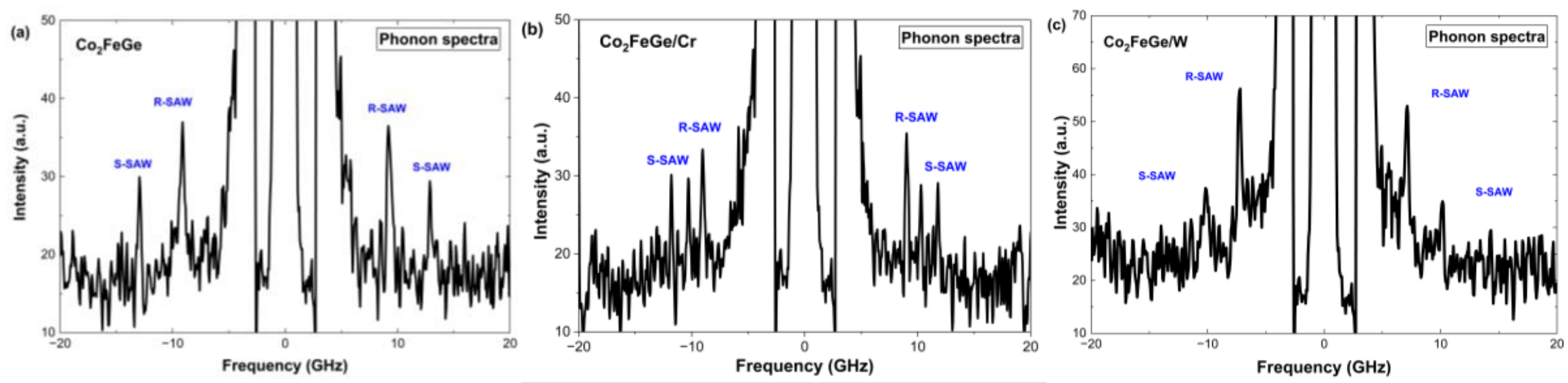

*Fig. 7: BLS spectra of phonon for the samples S1 (a), S2 (b), and S3 (c), $q = 13.1\ \mu m^{-1}$. The R-SAW and S-SAW surface acoustic modes are visible.*

The phonon dispersion relation obtained from the BLS measurements is shown in Fig. 8(a), (b), and (c). The experimentally measured dispersion curves show good agreement with the results obtained from the finite element method (FEM) simulations for all three samples with different buffer layers. The introduction of buffer layers shifts the acoustic branches to lower frequencies compared with the buffer-free sample. This behavior demonstrates that the buffer layer significantly modifies the effective surface-acoustic-wave velocity of the multilayer structure. The observed frequency reduction originates primarily from the increased effective mass loading and the altered acoustic impedance introduced by the metallic buffer layer. Because W possesses a substantially higher mass density and a lower transverse acoustic velocity than MgO, its influence on the effective elastic response is considerably stronger than that of Cr, resulting in the largest downshift of the acoustic branches.

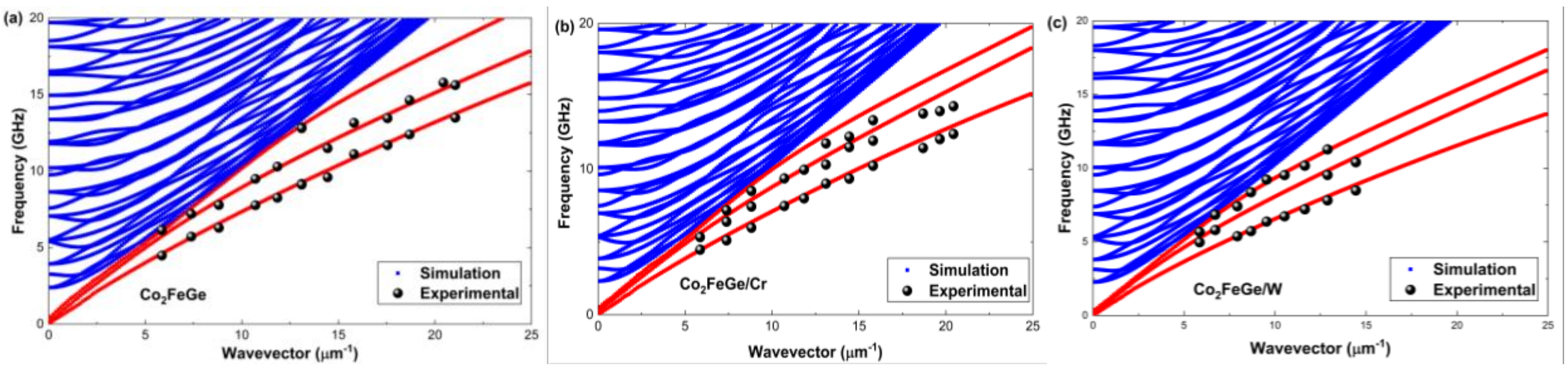

*Fig. 8: Phonon dispersion of sample S1 (a), S2 (b), and S3 (c). In phonon dispersion, black spheres correspond to the experimental data with error and the continuous lines correspond to the simulated data. The first red line corresponds to R-SAW, and the next two red lines correspond to S-SAWs. The blue region represents the bulk phonons (sound cone).*

A systematic comparison of buffer-free, Cr-buffered, and W-buffered $Co_2FeGe/MgO(001)$ heterostructures reveals that buffer-layer modification affects both the effective elastic response and the dynamic magnetic response of the films, but with contrasting trends for acoustic and spin-wave excitations.

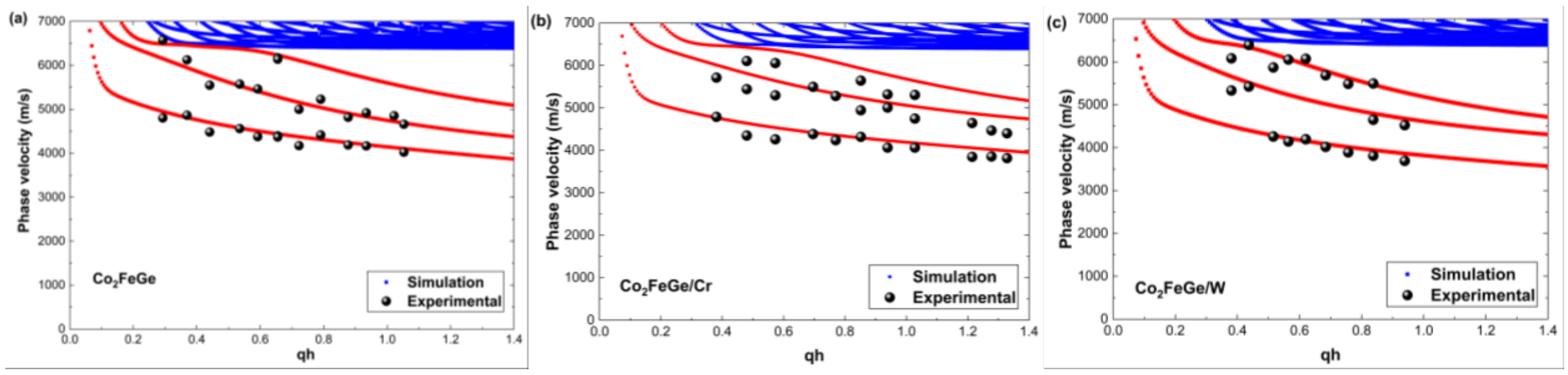

*Fig. 9: Phase velocity of the surface acoustic waves in samples S1, S2, and S3. Black symbols denote the experimental BLS data, while solid lines represent the FEM-simulated acoustic branches*

The phonon spectra clearly exhibit surface acoustic wave modes, including the R-SAW and the S-SAW. The appearance of the Sezawa mode is typical in layered structures and originates from the acoustic velocity contrast between the thin film and the substrate, which results in guided acoustic modes localized near the film/substrate region [48].

The highest frequencies are observed for the buffer-free sample (S1), followed by the Cr-buffered sample (S2), while the W-buffered sample (S3) exhibits the lowest frequencies. Compared to S1, the R-SAW frequency decreases by approximately 3% in S2 and by approximately 16% in S3, indicating a strong buffer-layer influence on the effective acoustic response. The experimentally measured phonon dispersion is in good agreement with the FEM simulations, supporting the use of the model to interpret the observed SAW branches [29,48].

Figure 9 presents the phase velocities of the surface acoustic waves for samples S1, S2, and S3. A clear buffer-layer dependence is evident in the measured phase velocities. The buffer-free $Co_2FeGe$ sample (S1) exhibits the highest phase velocities, while the W-buffered sample (S3) shows the lowest values over the entire investigated wavevector range. The Cr-buffered sample (S2) displays an intermediate behavior. This trend indicates that the introduction of a buffer layer modifies the effective elastic properties of the heterostructure and consequently changes the effective SAW phase velocity.

The representation of phase velocity as a function of *qh* is particularly useful because it relates the acoustic wavelength to the effective thickness of the near-surface multilayer. Such dimensionless thickness parameters, commonly expressed as h/λ or *qh*, are widely used in layered SAW systems because the number of supported Rayleigh/Sezawa-type modes and their phase velocities strongly depend on the acoustic wavelength relative to the film thickness [49,50,51]. In this range, the acoustic displacement field is sensitive not only to the $Co_2FeGe$ layer but also to the buffer layer and the upper part of the MgO substrate. Therefore, the observed velocity reduction should not be interpreted as a simple density effect alone. Instead, it reflects the combined influence of mass loading, elastic constants, acoustic impedance contrast, interfacial stiffness, and the localization of the Rayleigh- and Sezawa-type modes in the layered structure [52].

The stronger downshift observed for the W-buffered sample can be rationalized by the combined effects of mass loading and acoustic impedance mismatch. Because W has a much higher mass density than Cr and $Co_2FeGe$, its inclusion increases the effective mass density of the near-surface multilayer, consistent with the known sensitivity of the Rayleigh and Sezawa SAW modes to mass loading in layered structures [53]. Although W also has a high elastic stiffness, the net effect in the present geometry is a reduction of the effective SAW phase velocity. The larger acoustic impedance mismatch introduced by W can also modify the confinement and localization of Rayleigh- and Sezawa-type modes, whose displacement profiles in layered systems strongly depend on the layer/substrate acoustic contrast and thickness-to-wavelength ratio [54].

Although DE and PSSW modes have previously been reported in $Co_2FeGe$/MgO thin films, the present work employs these excitations as sensitive probes of buffer-layer-induced modifications to the dynamic magnetic response. By correlating the spin-wave results with the simultaneously measured surface-acoustic-wave dispersion, the influence of Cr and W buffer layers on both magnetic and elastic excitations can be directly evaluated [55].

The observation of multiple PSSW modes provides additional sensitivity to the magnetic properties of the films because their frequencies depend on the exchange stiffness, magnetic anisotropy, and interfacial spin-pinning conditions. The differences in PSSW frequencies observed between the investigated samples therefore, suggest that the buffer layers modify the magnetic boundary conditions of the $Co_2FeGe$ films.

The Cr and W buffer layers can be interpreted as two physically distinct interface modifications. Cr buffer layers have previously been used in Co-based Heusler films to promote epitaxial growth on MgO and to modify structural ordering, damping, and magnetic anisotropy [56,57]. In the present samples, the Cr buffer produces only a moderate acoustic downshift, but it may still modify the spin-wave spectra through changes in strain, ordering, or magnetic boundary conditions. In contrast, W has a much stronger influence on the acoustic branches because of its high mass density and acoustic impedance, while its heavy-metal character may also contribute to modified magnetic boundary conditions. However, possible W-related spin–orbit effects cannot be isolated from the present data without additional damping, spin-pumping, or nonreciprocity analysis [17, 58,59].

Table 3: Group velocities ($v_g$) for SW for different $Co_2FeGe$-based samples S1, S2, and S3.

| Sample | vg (m/s) |
|---|---|
| S1 | 1534 ± 126 |
| S2 | 2055 ± 120 |
| S3 | 2073 ± 135 |

The increase of approximately 34% in the estimated DE-mode group velocity, from 1534 ± 126 m/s in S1 to about $2.06 \times 10^3$ m/s in the buffered samples, suggests that the metallic buffer layers modify the dynamic magnetic response of the $Co_2FeGe$ film. The fact that the Cr- and W-buffered samples show comparable group velocities within uncertainty, despite their markedly different acoustic downshifts, indicates that the acoustic and magnetic responses are governed by different physical mechanisms.

The markedly different acoustic responses of the Cr- and W-buffered samples contrast with their nearly identical DE-mode group velocities. This indicates that the elastic and magnetic responses are governed by different interfacial mechanisms. Whereas the SAW dispersion is primarily controlled by mass loading, elastic constants, and acoustic impedance, the spin-wave dynamics are more strongly influenced by magnetic boundary conditions, exchange interactions, and interfacial anisotropy. Consequently, buffer-layer engineering enables partially independent tuning of acoustic and magnetic excitations, which is advantageous for the design of hybrid spin-acoustic devices.

It should be emphasized that the present measurements do not constitute a direct determination of magnon-phonon coupling. No anticrossing between acoustic and spin-wave branches, controlled SAW excitation, or coupling constant is extracted here. Instead, the present work establishes how Cr and W buffer layers modify the acoustic and spin-wave spectra within the same $Co_2FeGe$/MgO heterostructures. These results establish buffer-layer engineering as an effective strategy for independently tailoring acoustic and magnetic excitations in $Co_2FeGe$ heterostructures and provide a foundation for future investigations of coherent magnon–phonon coupling [60].

## 5. CONCLUSION

In summary, this study demonstrates that buffer-layer engineering is a powerful tool for the selective modification of dynamic excitations in $Co_2FeGe$ Heusler thin films. By incorporating Cr and W buffer layers, we successfully tuned the surface acoustic wave dispersion, achieving a significant frequency reduction of up to 16% in the W-buffered heterostructures. Finite-element simulations confirmed that this acoustic downshift is primarily driven by mass loading and shifts in effective acoustic impedance.

Parallel magnetodynamic investigations via BLS and FMR revealed that while buffer layers dramatically alter the elastic response, they also significantly enhance the spin-wave group velocity, increasing it by approximately 34% compared to buffer-free films. Crucially, the differing trends between the acoustic and magnetic shifts indicate that these properties are governed by distinct interfacial mechanisms.

These results establish that the elastic and magnetic properties of Heusler-based heterostructures can be tailored largely independent of one another. This work provides experimental and theoretical foundation for the design of future hybrid magnonic devices, where the precise pre-optimization of individual excitation spectra is a prerequisite for achieving efficient, coherent magnon-phonon coupling.

## 6. Acknowledgments

This work was supported by the Polish National Science Centre under grant no: UMO-2020/37/B/ST3/03936. Portuguese team acknowledges financial support from FCT – Portuguese Foundation for Science and Technology through the projects LA/P/0095/2020 (LaPMET), UIDB/04968/2025, and from FEDER – European Regional Development Fund through the project 17142|COMPETE2030-FEDER-00854500 (SynRoLoD). This work was partially supported by Spanish Ministerio de Economía y Competitividad through project PID2020-112914RB-I00, and from regional Gobierno de Aragón through project E28 20R including FEDER funding.

7. Data Availability

The authors declare that all data generated or analyzed during this study are included in this published article.

8. Author Contributions

A.T. planned the studies, A.V.A. carried out the BLS measurements, fitting, analysis and wrote the first version of the manuscript. A.V., P.S., and P.A.A. prepared the samples. A.V., B.P. and J.P.A. carried out XRD measurements. S.B and G.N.K. carried out VNA-FMR measurements and fitting. A.T., A.V.A, A.V. and G.N.K. prepared the final version of the manuscript. All authors discussed the results and contributed to writing and improving the quality of the manuscript.